# People, IT, and Structuration (PIS): An Integrative Theoretical Framework for Management Information Systems

Wei Huang[1], Xiaofang Cai[1]*, Qiaozhen Guo[1], Xiaosong Wu[2], Xin Tang[3]

[1] Southern University of Science and Technology, Shenzhen, China

* Corresponding author. Email: caixf@mail.sustech.edu.cn

## Abstract

The Management Information Systems (MIS) discipline has long grappled with how to theorize the complex, mutually constitutive relationships among people, information technology, and organizational structures. Decades of research have produced influential but fragmented theoretical streams—from socio-technical systems theory to technology acceptance models, from adaptive structuration theory to sociomateriality—each illuminating important facets while leaving integrative questions unresolved. This paper proposes the **People–IT–Structuration (PIS)** framework as a unifying theoretical lens that synthesizes these streams. Drawing on Giddens' structuration theory, we conceptualize People (P), Information Technology (I), and Structure (S) not as independent variables but as mutually constitutive elements engaged in ongoing structuration processes. We trace the intellectual history of MIS theorizing to demonstrate how PIS resolves persistent tensions in the field: between technological and social determinism, between variance and process approaches, and between micro-level interaction and macro-level institutional dynamics. We develop a set of formal propositions articulating the mechanisms through which P, I, and S co-evolve, and extend the framework to address contemporary phenomena including artificial intelligence, algorithmic management, and human–AI collaboration. The PIS framework offers both a retrospective lens for understanding the discipline's theoretical evolution and a prospective tool for guiding research in the AI era.

**Acknowledgements**
*The authors sincerely appreciate all the support and insights that contributed to the development and completion of this theoretical study. We thank Professor Richard Watson and Professor K. K. Wei for their in-depth discussions, and collective efforts throughout the research process. Our joint deliberations on structuration theory, socio-technical perspectives, and contemporary AI-driven organizational phenomena greatly facilitated the construction and refinement of the People–IT–Structuration (PIS) integrative framework.*

# 1. Introduction

What is the fundamental subject matter of Management Information Systems (MIS) research? This question, deceptively simple, has generated some of the most consequential debates in the discipline's history (Benbasat & Zmud, 2003; Orlikowski & Iacono, 2001). At its core, MIS seeks to understand the intersection of people, technology, and organizations—yet the field has struggled to develop theoretical frameworks that treat these three elements as genuinely co-constitutive rather than as independent variables in causal models.

The theoretical landscape of MIS has evolved through distinct paradigmatic waves. The socio-technical systems (STS) tradition, originating in the Tavistock Institute's seminal studies (Trist & Bamforth, 1951; Emery & Trist, 1965), established the foundational insight that technical and social subsystems must be jointly optimized. The technology acceptance tradition, epitomized by the Technology Acceptance Model (TAM) (Davis, 1989) and its successors (Venkatesh et al., 2003), brought methodological rigor to understanding individual adoption but at the cost of treating technology as an exogenous stimulus and organizational context as a moderating variable. The structuration tradition, beginning with Orlikowski (1992) and DeSanctis and Poole (1994), introduced Giddens' (1984) social theory to explain how technologies are shaped by and shape organizational practices—but these applications remained partial, typically foregrounding either the technology–practice nexus or the technology–institution nexus without fully integrating the triadic relationship.

Meanwhile, the "IS artifact" debate (Orlikowski & Iacono, 2001; Benbasat & Zmud, 2003; Weber, 2003) exposed a deeper identity crisis: the discipline risked either dissolving its distinctive focus on IT into general organizational theory or reducing complex socio-technical phenomena to simplistic input–output models. The sociomateriality turn (Orlikowski, 2007; Leonardi, 2012) attempted resolution through ontological fusion of the social and material, but this move generated its own controversies about whether collapsing analytical distinctions sacrifices explanatory precision (Mutch, 2013).

Today, the emergence of artificial intelligence—systems that exhibit agency-like properties, make autonomous decisions, and reshape organizational structures in real time—renders these theoretical tensions more consequential than ever. AI is not merely a new technology to be "accepted" or "adopted." It fundamentally reconfigures the relationships among human actors, technological artifacts, and institutional arrangements in ways that existing frameworks, developed for an era of passive tools, cannot adequately capture.

This paper proposes the **People–IT–Structuration (PIS)** framework as an integrative theoretical response. PIS synthesizes the insights of prior MIS theoretical streams while resolving their limitations. Its core claim is straightforward yet powerful: *People (P), Information Technology (I), and Structure (S) are three analytically distinct but ontologically inseparable elements that mutually constitute each other through ongoing processes of structuration.* This triadic structuration—not pairwise interactions between any two elements—is the fundamental unit of analysis for MIS research.

The stakes of this theoretical project extend beyond academic debate. As organizations worldwide invest trillions of dollars in AI implementation, the frameworks guiding these investments shape real outcomes for workers, managers, and societies. Frameworks that treat AI as an exogenous variable to be "adopted" (Wave 2 thinking) lead to implementation strategies focused on user acceptance and change management. Frameworks that emphasize joint optimization of social and technical systems (Wave 1 thinking) lead to participatory design approaches. Frameworks that attend to appropriation dynamics (Wave 3 thinking) lead to attention to emergent use patterns. Each is valuable but partial. PIS argues that effective AI implementation requires simultaneous attention to all three elements—and that failure to do so explains a significant portion of the organizational dysfunction, resistance, and unintended consequences that characterize contemporary AI deployments.

The contributions of this paper are threefold. First, we provide a systematic intellectual history of MIS theorizing, revealing how each major stream illuminated one or two sides of the P–I–S triangle while neglecting the third. Second, we formally articulate the PIS framework, specifying its ontological commitments, core constructs, and structuration mechanisms. Third, we demonstrate the framework's generative power by applying it to contemporary AI-era phenomena, showing how PIS offers analytical purchase where existing frameworks fall short.

The remainder of this paper proceeds as follows. Section 2 traces the historical evolution of MIS theorizing through four waves. Section 3 develops the PIS framework's theoretical foundations. Section 4 presents formal propositions. Section 5 extends PIS to the AI era. Section 6 discusses implications for research and practice, and Section 7 concludes.

## 2. Historical Evolution of MIS Theorizing: Four Waves

To understand why an integrative framework is needed, we must first trace how MIS theorizing has evolved—and what each wave left unresolved. We identify four major theoretical waves, each advancing understanding while creating characteristic blind spots (see Table 1).

### 2.1 Wave 1: Socio-Technical Systems (1950s–1970s)

The intellectual roots of MIS lie in socio-technical systems theory, developed at London's Tavistock Institute of Human Relations. Trist and Bamforth's (1951) landmark study of British coal mining demonstrated that introducing new technology (longwall mining methods) disrupted established social systems, producing outcomes that neither technical efficiency models nor social analyses alone could predict. The key insight was *joint optimization*: because technical and social subsystems are interdependent, optimizing one at the expense of the other produces suboptimal system performance.

Emery and Trist (1965) extended this thinking to organizational environments, establishing that socio-technical systems are open systems embedded in turbulent environments requiring continuous adaptation. Mumford's (1983) ETHICS methodology operationalized STS principles for information systems design, emphasizing participatory approaches that attend to both technical requirements and human needs.

The STS tradition established two enduring principles for MIS: (1) technology and social organization are interdependent, and (2) effective system design requires attending to both simultaneously. However, STS operated with a relatively static conception of "the social" and "the technical" as two subsystems to be balanced, rather than as mutually constitutive elements in dynamic processes. The Tavistock researchers recognized interdependence but conceptualized it in systems-engineering terms—finding the right "fit" between subsystems—rather than as an ongoing process of mutual constitution. Structure—understood as institutional arrangements, power relations, and organizational rules—remained largely implicit, treated as the background context in which socio-technical optimization occurs rather than as an active element being continuously reproduced and transformed through technology use. As Trist (1981) later acknowledged, the STS framework struggled to account for the ways in which macro-level institutional environments shaped and were shaped by local socio-technical arrangements. The framework provided a powerful design heuristic but lacked a theoretical mechanism for explaining how socio-technical systems evolve dynamically over time.

## 2.2 Wave 2: Technology Acceptance and Adoption (1980s–2000s)

The second wave shifted focus from system design to individual-level technology use. Davis's (1989) Technology Acceptance Model, rooted in the Theory of Reasoned Action (Fishbein & Ajzen, 1975), proposed that technology use is determined primarily by perceived usefulness and perceived ease of use. TAM's parsimony and predictive power generated an enormous research stream, culminating in the Unified Theory of Acceptance and Use of Technology (UTAUT), which integrated eight competing models into a single framework identifying four key determinants of behavioral intention (Venkatesh et al., 2003).

This wave brought methodological sophistication to MIS, establishing robust measurement instruments and statistical techniques. Yet its theoretical limitations became increasingly apparent. TAM and its variants treat technology as an exogenous variable—a stimulus to which individuals respond—rather than as something shaped through use. Organizational structures enter only as moderating variables (e.g., voluntariness, social influence) rather than as constitutive elements. The framework is fundamentally unidirectional: technology influences people's perceptions, which determine use. The recursive loops through which use transforms both the technology's meaning and the organizational context are invisible.

As Benbasat and Barki (2007) observed, the TAM research stream, despite (or because of) its popularity, contributed to a "theoretical straightjacket" that constrained MIS theorizing. By reducing the complex entanglement of people, technology, and organizations to individual adoption decisions, TAM effectively excised the "M" (management) and the "S" (systems) from MIS.

The limitations of this wave become especially stark in the AI era. TAM assumes that technology is a stable object whose properties can be perceived and evaluated prior to adoption. But AI systems that learn, adapt, and generate novel outputs present a moving target—their "usefulness" and "ease of use" are not fixed properties but emergent outcomes of ongoing interaction. Moreover, TAM's individualistic focus cannot capture the collective, organizational dynamics through which AI reshapes work practices, authority relations, and institutional arrangements. The wave's core contribution—establishing that individual perceptions mediate technology use—remains valid but radically incomplete as a theoretical foundation for MIS.

## 2.3 Wave 3: Structuration and Practice (1990s–2010s)

The third wave introduced social theory to MIS, particularly Giddens' (1984) structuration theory. Giddens' central concept—the *duality of structure*, in which structures are both the medium and outcome of social action—offered a theoretical vocabulary for transcending the determinism debates that plagued earlier waves.

Orlikowski (1992) was among the first to apply structuration theory to technology, proposing a "duality of technology" in which technology is physically constructed by human actors in a given social context and socially constructed through the meanings attributed to it. DeSanctis and Poole's (1994) Adaptive Structuration Theory (AST) provided a more operational framework, distinguishing between the structural features embedded in technology ("spirit" and structural features) and the ways groups appropriate these features in practice.

Orlikowski (2000) later moved beyond the duality model to a practice lens emphasizing "technologies-in-practice"—the recurrent, situated engagement with technology through which people accomplish their work. This shift from structures-in-technology to structures enacted through technology use represented a significant theoretical advance, foregrounding the emergent, processual nature of technology–organization relationships.

However, structuration applications in IS tended to privilege certain pairwise relationships. Orlikowski's work primarily theorized the People–IT nexus (how human practices constitute technology-in-use), while AST focused on the IT–Structure nexus (how technological features are appropriated within group structures). The full triadic relationship—in which People, IT, and Structure simultaneously constitute each other—was implied but never systematically developed. Moreover, the "structure" concept remained underspecified, sometimes referring to institutional arrangements, sometimes to group norms, and sometimes to technology's structural features.

Jones and Karsten's (2008) comprehensive review of structuration theory applications in IS confirmed this assessment. Analyzing 331 articles that referenced Giddens' work, they found that most applications were partial and selective, drawing on particular concepts (especially the duality of structure) without engaging the full apparatus of structuration theory. They noted a tendency to "cherry-pick" from Giddens, using structuration language to describe dynamics that could be captured by simpler frameworks, while leaving the more challenging aspects—the interplay of signification, domination, and legitimation; the relationship between system integration and social integration; the role of unintended consequences in institutional reproduction—largely untouched. This selective appropriation left the integrative potential of structuration theory for IS unrealized.

## 2.4 Wave 4: The IS Artifact Debate and Sociomateriality (2000s–2020s)

The fourth wave was precipitated by growing anxiety about the discipline's identity. Orlikowski and Iacono (2001) provocatively argued that MIS research had failed to adequately theorize its core subject: the IT artifact. Reviewing 188 articles in *Information Systems Research*, they found that most treated IT as absent (studied context without technology), black-boxed (present but unexamined), abstracted (reduced to a variable), or nominalized (named but not analyzed).

Benbasat and Zmud (2003) responded by proposing that MIS should focus on "the IT artifact and its immediate nomological net"—the constructs and relationships directly surrounding technology use in organizational contexts. Weber (2003) countered that this boundary was either too narrow (excluding important phenomena) or too vague (providing insufficient guidance).

The sociomateriality perspective (Orlikowski, 2007; Orlikowski & Scott, 2008) attempted a more radical resolution, arguing that the social and the material are "constitutively entangled"—there is no social that is not also material, and no material that is not also social. While theoretically provocative, this ontological move drew criticism for collapsing analytical distinctions needed for empirical research (Leonardi, 2012; Mutch, 2013). Leonardi (2012) advocated instead for a "sociomaterial" perspective that maintains analytical separation while studying empirical entanglement.

This wave exposed a fundamental tension that remains unresolved: MIS needs frameworks that acknowledge the deep interdependence of people, technology, and structure without either (a) collapsing them into an undifferentiated mass or (b) treating them as independent variables in causal models. It is precisely this tension that the PIS framework addresses.

**Table 1. Four Waves of MIS Theorizing: Contributions and Blind Spots**

| Wave | Key Works | Core Contribution | What Was Left Unresolved |
|---|---|---|---|
| 1. Socio-Technical Systems | Trist & Bamforth (1951); Emery & Trist (1965); Mumford (1983) | Joint optimization of social and technical subsystems | Structure treated as static background; no mechanism for dynamic co-evolution |
| 2. Technology Acceptance | Davis (1989); Venkatesh et al. (2003) | Rigorous models of individual adoption | Technology as exogenous; organizational context as moderator; no recursion |
| 3. Structuration & Practice | Orlikowski (1992, 2000); DeSanctis & Poole (1994) | Duality of technology; appropriation; practice lens | Pairwise focus (P–I or I–S); "structure" underspecified; triadic dynamics implicit |
| 4. IS Artifact & Sociomateriality | Orlikowski & Iacono (2001); Leonardi (2012) | Attention to IT artifact; constitutive entanglement | Tension between ontological fusion and analytical distinction unresolved |

# 3. The PIS Framework: Theoretical Foundations

## 3.1 From Pairwise to Triadic: The Integrative Gap

The historical review reveals a consistent pattern: each theoretical wave foregrounded one or two elements of the People–IT–Structure relationship while backgrounding the third. STS attended to People and IT (as social and technical subsystems) but left Structure implicit. TAM focused on the People–IT relationship (perceptions driving use) while treating Structure as an external moderator. Structuration approaches oscillated between People–IT (practice lens) and IT–Structure (AST) without fully integrating all three. Sociomateriality attempted integration through ontological fusion but sacrificed the analytical distinctions needed for empirical research.

What is missing is a framework that simultaneously:
1. Maintains analytical distinction among People, IT, and Structure as three theoretically meaningful elements;
2. Theorizes their mutual constitution—each element is continuously produced and reproduced through its interaction with the other two;
3. Specifies the mechanism of this mutual constitution as structuration—the recursive process through which action and structure co-evolve over time;
4. Operates across levels of analysis, from micro-level human–technology interaction to macro-level institutional transformation.

The PIS framework is designed to fill this integrative gap.

## 3.2 Core Constructs

PIS is built on three core constructs, each defined to be analytically distinct while acknowledging their ontological interdependence.

**People (P).** "People" encompasses human actors with agency—the capacity to act intentionally, reflexively monitor their actions, and rationalize their conduct (Giddens, 1984). In the PIS framework, People includes individual users, groups, managers, designers, and other stakeholders who engage with information technology within organizational contexts. Crucially, People are not passive recipients of technology or institutional pressures; they are knowledgeable agents who draw upon interpretive schemes, norms, and resources in their interactions with IT and organizational structures. People's agency manifests in three modes relevant to PIS: *interpretive agency* (the capacity to assign meaning to technologies and structures, constructing shared understandings that shape collective action); *practical agency* (the capacity to use technologies skillfully in everyday work, developing routines and workarounds that may diverge from designed functionality); and *political agency* (the capacity to mobilize resources, exercise power, and negotiate competing interests in shaping technology design and organizational arrangements).

**Information Technology (I).** "Information Technology" refers to the material and digital artifacts—hardware, software, algorithms, data, networks, and platforms—that mediate, augment, or automate human activity within organizational settings. Following Orlikowski and Iacono (2001), we insist on theorizing IT rather than black-boxing it. Following Leonardi (2012), we maintain that IT possesses material properties (affordances and constraints) that are analytically distinguishable from—though empirically entangled with—the social practices in which it is embedded. In the PIS framework, IT is characterized by: *material affordances* (the action possibilities that technology's material properties make available to users within particular contexts); *inscribed structures* (the rules, assumptions, and logics embedded in technology through design processes, reflecting designers' understandings of work and organization); and *generativity* (the capacity of technology to produce outcomes not fully anticipated by designers or users, particularly salient in AI systems).

**Structure (S).** "Structure" denotes the institutional arrangements, organizational rules, cultural norms, power relations, and resource distributions that both enable and constrain human action (Giddens, 1984). In Giddens' formulation, structure comprises three dimensions: signification (meaning), domination (power), and legitimation (norms). In MIS contexts, Structure manifests as: *organizational structures* (hierarchies, roles, policies, workflows, and governance mechanisms); *institutional logics* (the broader cultural and cognitive frameworks that define appropriate action in particular domains); and *resource configurations* (the distribution of authoritative resources and allocative resources within and across organizational boundaries). Critically, Structure in PIS is not a static "context" variable. It is continuously produced, reproduced, and potentially transformed through the actions of People using IT. This is the essence of structuration.

## 3.3 The Structuration Mechanism: Triadic Co-Constitution

The theoretical engine of PIS is *triadic structuration*: the ongoing, recursive process through which People, IT, and Structure mutually constitute each other. Unlike prior applications of structuration theory in IS that focused on pairwise duality (e.g., Orlikowski's duality of technology), PIS insists that the unit of structuration is always triadic.

Consider an example: when a team adopts a new AI-powered project management tool. People (project managers, team members) interact with the IT (the AI tool's algorithms, interfaces, recommendations) within an existing Structure (organizational hierarchy, project governance norms, performance metrics). Through use, People develop new practices (practical agency), reinterpret their roles (interpretive agency), and negotiate new responsibilities (political agency). The IT, through its inscribed structures and generative capabilities, enables certain practices while constraining others, and its algorithmic outputs may shift over time as it learns from use patterns. The Structure—governance arrangements, authority relations, norms of accountability—is simultaneously reproduced (to the extent that existing patterns persist) and transformed (as new practices, empowered by the technology, reshape organizational expectations).

This triadic process operates through three interrelated structuration circuits:

**Circuit 1: P↔I (Human–Technology Structuration).** People appropriate IT through use, developing "technologies-in-practice" (Orlikowski, 2000) that may conform to or diverge from designed functionality. Simultaneously, IT's material affordances and constraints shape people's capabilities, perceptions, and identities. This circuit captures the insights of TAM (technology shaping perceptions) and the practice lens (people shaping technology-in-use) but treats them as a single recursive process rather than separate phenomena.

**Circuit 2: P↔S (Human–Structure Structuration).** People draw upon structural resources and rules to guide their action while simultaneously reproducing or transforming those structures through their actions. This is the core of Giddens' duality of structure as applied to organizational life. In MIS contexts, this circuit is mediated by technology—people engage with structures partly *through* their technology use—but analytically, the P↔S circuit captures dynamics of power, legitimation, and meaning-making that are not reducible to technology.

**Circuit 3: I↔S (Technology–Structure Structuration).** Technology embodies structural properties through its design (inscribed structures) and, through its deployment and use, reshapes organizational structures. Enterprise systems enforce workflow standardization; AI algorithms reconstruct decision-making hierarchies; platforms create new organizational forms. Reciprocally, organizational structures shape technology trajectories through procurement decisions, customization choices, and governance frameworks. This circuit captures insights from institutional theory (Fountain, 2001) and the "mutual shaping" tradition while specifying the structuration mechanism.

**The Triadic Nexus.** The critical insight of PIS is that these three circuits do not operate independently. They are coupled: changes in any one circuit propagate through the others. When AI reshapes the I↔S circuit (e.g., algorithmic management replacing middle-management oversight), this simultaneously transforms the P↔I circuit (people develop new relationships with AI systems) and the P↔S circuit (authority relations and career structures shift). It is this triadic coupling that generates the emergent, often unpredictable dynamics that characterize contemporary IS phenomena.

## 3.4 Illustrative Application: Enterprise AI Implementation

To clarify how triadic structuration operates in practice, consider a concrete scenario that illuminates the framework's analytical power. A large hospital system implements an AI-powered clinical decision support system (CDSS) to assist physicians in diagnosing complex cases.

In a TAM analysis, the key questions would be: Do physicians perceive the CDSS as useful? As easy to use? These perceptions would predict adoption. In an STS analysis, the focus would be on jointly optimizing the technical system (algorithm accuracy, interface design) and the social system (clinical workflows, team communication). In an AST analysis, the focus would be on how clinical teams appropriate the CDSS's structural features (recommendations, confidence scores, explanation interfaces) faithfully or unfaithfully.

A PIS analysis encompasses and transcends all three. It examines the *simultaneous, recursive* dynamics across all three circuits:

In the P↔I circuit, physicians develop "technologies-in-practice"—some integrate AI recommendations into their diagnostic reasoning, others use them as a second opinion to confirm existing judgments, still others develop workarounds to satisfy institutional requirements for AI consultation while maintaining their own diagnostic authority. These divergent appropriation patterns are not noise to be controlled but central phenomena to be explained.

In the I↔S circuit, the CDSS embodies particular assumptions about diagnostic workflow (sequential hypothesis testing), authority relations (AI as advisor, physician as decision-maker), and accountability structures (the physician remains legally responsible). Through deployment, the system reshapes these very structures: clinical pathways are reorganized around AI consultations, new roles emerge (AI coordinators), and quality metrics are redefined to incorporate AI-concordance rates.

In the P↔S circuit, physicians' professional identity and authority are simultaneously threatened and reinforced. Those who effectively integrate AI may gain status and efficiency; those who resist may face institutional pressure. Meanwhile, physicians collectively negotiate new norms about when AI consultation is appropriate, what constitutes "good" AI-augmented clinical judgment, and who bears responsibility when AI-informed decisions go wrong.

Crucially, these three circuits are coupled. The way physicians appropriate the CDSS (P↔I) depends on organizational governance structures (S); the structural changes triggered by AI deployment (I↔S) reshape physicians' agency and identity (P↔S); and physicians' collective negotiation of new professional norms (P↔S) feeds back into how they interact with the technology (P↔I) and what organizational adaptations are possible (I↔S). No pairwise analysis captures these coupled dynamics; only triadic structuration does.

## 3.5 Ontological and Epistemological Commitments

The PIS framework adopts a *critical realist* ontological position (Bhaskar, 1975), consistent with Giddens' own philosophical commitments and with calls for ontological pluralism in IS research (Mingers, 2001). This position holds that:

1. People, IT, and Structure exist as analytically distinct entities with real properties (contra strong sociomateriality's ontological fusion);

2. These properties are emergent—they arise from interaction and cannot be reduced to any single element (contra technological or social determinism);

3. Social structures, while real, exist only insofar as they are instantiated in practice (the structuration principle);

4. Knowledge of P–I–S dynamics is always partial, perspectival, and historically situated (epistemological humility).

This positioning allows PIS to navigate between the Scylla of ontological fusion (losing analytical traction) and the Charybdis of variable-based decomposition (losing systemic understanding).

# 4. Propositions

We now articulate a set of formal propositions that specify the mechanisms of triadic structuration. These propositions are intended to be generative—guiding research questions and empirical investigation—rather than directly testable hypotheses.

## 4.1 Foundational Propositions

> **Proposition 1 (Triadic Inseparability).** *Any information systems phenomenon is constituted by the simultaneous interaction of People, IT, and Structure. Research that examines only one or two elements will produce systematically incomplete explanations.*

This proposition does not merely claim that context matters (a trivial observation). It claims that P, I, and S are *constitutive* of each other—each element's properties and behaviors are partly produced by the other two. A technology's "usefulness" is not an intrinsic property but an emergent outcome of how People appropriate it within particular Structural contexts. Similarly, an organizational structure's "rigidity" or "flexibility" depends partly on the technologies that instantiate and enforce (or enable circumvention of) its rules.

> **Proposition 2 (Recursive Constitution).** *People, IT, and Structure are simultaneously the medium and outcome of structuration processes. Each element enables and constrains the others while being continuously reproduced and transformed through their interaction.*

This extends Giddens' duality of structure to the triadic case. Structures provide rules and resources that People draw upon in their interactions with IT; these interactions reproduce (and potentially transform) the structures; the transformed structures then reshape subsequent interactions. The recursion operates across all three circuits simultaneously.

## 4.2 Circuit-Specific Propositions

> **Proposition 3 (Appropriation Divergence).** *The structuration outcomes of technology use will systematically diverge from the intentions of technology designers to the extent that (a) People exercise interpretive and practical agency in appropriation, and (b) the receiving Structural context differs from the context assumed in design.*

This proposition integrates insights from AST's "unfaithful appropriation" concept (DeSanctis & Poole, 1994) with institutional theory's recognition that technologies are interpreted through local institutional logics. It explains why identical technologies produce different outcomes in different organizations—not as a residual to be controlled but as a fundamental feature of triadic structuration.

> **Proposition 4 (Structural Inscription and Drift).** *Information technologies inscribe structural assumptions at design time, but these inscriptions are subject to interpretive drift as People appropriate technologies within evolving Structural contexts. The rate and direction of drift depends on the technology's interpretive flexibility and the structural context's institutional stability.*
>
> **Proposition 5 (Generative Coupling).** *Changes in any one structuration circuit (P↔I, P↔S, or I↔S) propagate to the other two circuits, generating emergent systemic effects that are not predictable from analysis of any single circuit alone.*

This is perhaps PIS's most distinctive proposition. It explains why technology implementations frequently produce "unintended consequences"—not because of poor planning, but because triadic coupling means that interventions in one circuit inevitably reverberate through the others in ways that linear models cannot capture.

## 4.3 Structural Propositions

> **Proposition 5b (Structural Reproduction Dominance).** *In the absence of significant disruption to any element, triadic structuration tends toward reproduction rather than transformation. Existing structures shape technology design and appropriation in ways that reinforce current arrangements, and established technologies constrain the structural changes that people can enact. This reproductive tendency explains the persistence of organizational patterns even in the face of nominally transformative technology implementations.*

This proposition addresses a puzzle frequently observed in IS research: why do organizations often reproduce existing power relations, workflows, and institutional arrangements even after implementing technologies designed to transform them? PIS explains this as a natural consequence of structuration's reproductive tendency—structures shape how technologies are designed, selected, and appropriated in ways that tend to reproduce existing arrangements. Transformation, when it occurs, typically requires disruption significant enough to destabilize the reproductive equilibrium across multiple circuits simultaneously.

## 4.4 Temporal Propositions

> **Proposition 6 (Structuration Tempo).** *The three structuration circuits operate at different temporal rhythms. P↔I structuration (interaction-level) operates at the fastest tempo; I↔S structuration (organizational-level) at an*

> *intermediate tempo; and the institutional dimensions of P↔S structuration at the slowest tempo. Misalignment among these tempos generates tensions that can either inhibit or catalyze system transformation.*
>
> **Proposition 7 (Punctuated Structuration).** *Triadic structuration alternates between periods of reproductive stability (in which the three circuits reinforce existing patterns) and episodes of transformative disruption (in which changes in one element—typically IT—destabilize established circuits and trigger widespread restructuration). The introduction of AI into organizational contexts represents such a punctuation.*

# 5. PIS in the AI Era

The emergence of artificial intelligence provides both the motivation for the PIS framework and its most compelling application domain. AI systems—particularly large language models, autonomous agents, and algorithmic decision-making systems—challenge existing IS frameworks in three fundamental ways that PIS is uniquely equipped to address.

## 5.1 AI as a Qualitative Shift in the IT Element

Prior IS frameworks were developed for an era in which IT was fundamentally a tool—a passive artifact that humans designed, deployed, and used. AI disrupts this assumption. Contemporary AI systems exhibit properties that blur the boundary between People and IT: *autonomous agency* (AI systems can make decisions, take actions, and produce outputs without direct human instruction); *adaptive learning* (AI systems modify their own behavior based on experience, making them "moving targets" for structuration processes); and *generative capability* (large language models can produce novel content—text, code, analysis, strategy—that was previously the exclusive domain of human agency).

Within PIS, these properties do not require abandoning the P–I–S distinction. Rather, they intensify and complicate the structuration circuits. AI's agency-like properties mean that the P↔I circuit now involves something closer to interaction between two agents rather than a human using a tool. The I↔S circuit becomes more dynamic as AI systems continuously reshape organizational processes without human mediation. And the P↔S circuit is transformed as AI-mediated structures may operate opaquely, creating new challenges for human agency and institutional accountability.

## 5.2 Reconceptualizing the Structuration Circuits for AI

**P↔I in the AI era: From use to collaboration.** When IT exhibits agency-like properties, the human–technology relationship shifts from "use" (a unidirectional concept) to "collaboration" (a bidirectional concept). Human–AI collaboration involves mutual adaptation: humans adjust their practices to AI capabilities and limitations, while AI systems (through learning) adjust to human preferences and organizational contexts. PIS frames this as an intensified structuration circuit in which both parties are active agents in constituting shared practices.

**I↔S in the AI era: Algorithmic structuration.** AI-powered systems increasingly function as *structural agents*—they do not merely reflect organizational structures but actively produce them. Algorithmic management systems assign tasks, evaluate performance, and enforce rules without human intervention, effectively encoding and enacting organizational structure in real time (Kellogg et al., 2020). PIS conceptualizes this as a new mode of I↔S structuration in which technology does not merely inscribe structural assumptions but dynamically generates and enforces structural arrangements.

**P↔S in the AI era: Mediated agency.** As AI systems intermediate between people and organizational structures, human agency becomes increasingly *mediated*. Workers interact with organizational structures partly through AI systems that filter, interpret, and sometimes override human judgment. This creates new challenges for the P↔S circuit: how do people exercise political agency when algorithmic structures are opaque? How are norms of accountability maintained when decisions emerge from human–AI collaboration rather than individual human judgment?

## 5.3 From Management Elimination to Management Transformation

A popular narrative surrounding AI asserts that artificial intelligence will render management obsolete—that algorithms will optimize organizational processes so effectively that human managerial judgment becomes unnecessary. PIS provides a theoretically grounded counter-narrative that is both more nuanced and more empirically defensible.

The elimination thesis treats management as a set of functions (planning, organizing, directing, controlling) that can be automated once AI reaches sufficient capability. This is essentially a technological determinist argument operating within the I↔S circuit: technology (I) directly reshapes organizational structure (S), with people (P) merely adapting to the new arrangement. PIS reveals the poverty of this analysis by insisting on the triadic nature of organizational structuration.

Management, in PIS terms, is not a function but a *structuration practice*—an ongoing process through which organizational actors (P) use available technologies (I) to produce, reproduce, and transform organizational structures (S). When AI enters the picture, it transforms this practice but does not eliminate it. The need for human judgment about *what is worth doing* (effectiveness), as distinct from *how to do it efficiently* (efficiency), remains irreducibly human because it involves interpretive, political, and normative dimensions that cannot be reduced to optimization.

This analysis yields an important insight about the digital economy more broadly. The characteristics of digital economic activity—speed, scale, network effects, data abundance—create new management challenges rather than eliminating the need for management. Organizations, enterprises, and governance structures do not disappear in the digital economy; they transform. PIS provides the theoretical vocabulary for studying this transformation as a structuration process rather than a replacement event.

## 5.4 PIS and the Division of Labor Between Humans and AI

The PIS framework offers a theoretical basis for one of the most pressing questions in contemporary organizations: how should labor be divided between humans and AI? Rather than treating this as a technical optimization problem (matching tasks to capabilities), PIS frames it as a structuration question: *What arrangements of P, I, and S produce sustainable, effective, and legitimate organizational structuration?*

This reframing suggests that the effectiveness of human–AI collaboration depends not only on task characteristics but on the triadic alignment among human agency (People), AI capabilities and limitations (IT), and organizational governance arrangements (Structure). An AI system that is technically capable of autonomous decision-making may produce poor outcomes if the organizational structure lacks accountability mechanisms for AI-mediated decisions, or if people lack the interpretive frameworks to understand and appropriately trust AI recommendations.

> **Proposition 8 (AI Structuration Alignment).** *The effectiveness and sustainability of human–AI organizational arrangements depend on triadic alignment: the degree to which People's agency capacities, IT's autonomous capabilities, and Structural governance mechanisms are mutually supportive. Misalignment among any pair—such as AI capability exceeding structural governance or human agency being undermined by algorithmic opacity—produces structuration tensions that manifest as organizational dysfunction.*
>
> **Proposition 9 (Dual Optimization Thesis).** *AI enables organizations to pursue both effectiveness (a People-centered quality requiring judgment, creativity, and contextual understanding) and efficiency (an IT-centered quality requiring speed, consistency, and scale) simultaneously, but only when organizational Structures evolve to support a division of labor that leverages human and AI comparative advantages rather than substituting one for the other.*

# 6. Discussion

## 6.1 Comparison with Existing Frameworks

To sharpen PIS's distinctive contributions, Table 2 compares it systematically with the major frameworks reviewed in Section 2 across seven analytical dimensions.

**Table 2. Systematic Comparison of MIS Theoretical Frameworks**

| Dimension | STS | TAM/UTAUT | AST | Sociomateriality | PIS |
|---|---|---|---|---|---|
| Unit of analysis | Work system | Individual user | Group appropriation | Socio-material assemblage | Triadic structuration episode |
| Role of IT | Subsystem to be optimized | Exogenous stimulus | Structural features for appropriation | Constitutively entangled with social | Analytically distinct; materially affordant; generative |
| Role of people | Subsystem to be optimized | Perceiver and adopter | Appropriating agents | Entangled with material | Agentic (interpretive, practical, political) |
| Role of structure | Implicit context | Moderating variable | Group structures | Fused with material | Co-constitutive; three Giddens dimensions |
| Causal logic | Joint optimization | Variance model | Appropriation process | Relational ontology | Recursive triadic structuration |
| Temporal scope | Design-time | Pre-adoption | Implementation | Ongoing practice | Multi-tempo; design through evolution |
| AI readiness | Low | Low | Medium | Medium | High |

The comparison reveals PIS's two distinctive advantages. First, PIS is the only framework that explicitly treats all three elements—People, IT, and Structure—as co-constitutive. Other frameworks either fuse elements (sociomateriality), omit one (TAM omits Structure; STS backgrounds it), or address only pairwise relationships (AST). Second, PIS's multi-tempo structuration mechanism makes it uniquely suited to studying AI phenomena, which involve rapid interaction-level adaptation (P↔I), medium-term organizational restructuring (I↔S), and slow institutional evolution (P↔S) operating simultaneously.

## 6.2 Theoretical Contributions

The PIS framework makes three primary contributions to MIS theory. First, it provides an *integrative synthesis* of the field's major theoretical streams. Rather than proposing yet another competing perspective, PIS shows how STS, technology acceptance, structuration, and sociomateriality each captured important aspects of a fundamentally triadic phenomenon. By making the P–I–S triangle explicit and specifying the structuration mechanism that connects its elements, PIS offers a meta-theoretical framework within which prior contributions can be understood as partial views.

Second, PIS resolves the *ontological tension* between sociomaterial fusion and variable-based decomposition that has preoccupied IS theory. By adopting a critical realist position, PIS maintains that P, I, and S are analytically distinct (enabling precise theorizing and empirical investigation) while insisting that they are ontologically co-constitutive (preventing reductive decomposition). The three structuration circuits provide the mechanism through which analytical distinction and ontological interdependence coexist.

Third, PIS provides a *generative theoretical platform* for studying AI-era phenomena. Unlike frameworks developed for passive tools, PIS can accommodate IT with agency-like properties, continuous organizational restructuring driven by algorithms, and the emergent dynamics of human–AI collaboration. The framework's propositions identify specific mechanisms (appropriation divergence, generative coupling, structuration tempo, AI alignment) that generate testable implications for empirical research.

## 6.3 Implications for Research

PIS implies several reorientations in MIS research practice:

*Unit of analysis.* PIS suggests that the appropriate unit of analysis for MIS research is not the individual (as in TAM), the technology (as in artifact-focused work), or the organization (as in institutional studies), but the *structuration episode*—a bounded process in which P, I, and S interact to produce, reproduce, or transform their mutual constitution. Research designs should be configured to capture triadic dynamics rather than pairwise relationships.

*Temporal design.* Given Proposition 6 (structuration tempo), PIS favors longitudinal research designs that can capture the different temporal rhythms of the three circuits. Cross-sectional studies inevitably capture snapshots of an ongoing process, missing the recursive dynamics that are central to PIS.

*Methodological pluralism.* Because the three circuits operate at different levels and tempos, PIS supports—indeed requires—methodological pluralism. Ethnographic and qualitative methods capture the micro-level dynamics of the P↔I circuit; organizational case studies capture the I↔S circuit; and institutional analysis captures the macro-level P↔S dynamics.

*AI-specific research agenda.* PIS opens specific research questions for the AI era that are not readily addressable by existing frameworks:

*Human–AI team composition.* How do different configurations of humans and AI agents within teams produce different structuration dynamics? McGrath's (1984) task typology, developed for human groups, may need fundamental extension when team members include AI agents with different agency profiles. PIS predicts that the composition of human–AI teams will reshape not only task performance (a P↔I question) but also authority structures, coordination norms, and accountability mechanisms (P↔S and I↔S questions).

*Algorithmic governance and legitimacy.* As AI systems increasingly enact organizational structures, questions of legitimacy become central. Under what conditions do people accept algorithmic structuration as legitimate? How do organizations develop governance mechanisms for AI-mediated structures? PIS's attention to the legitimation dimension of structure (following Giddens) provides theoretical tools for investigating these questions.

*AI and the transformation of management.* Perhaps the most far-reaching implication of PIS for the AI era concerns the nature of management itself. If AI can handle efficiency-oriented tasks (scheduling, monitoring, routine decision-making) while humans focus on effectiveness-oriented challenges (innovation, meaning-making, stakeholder alignment), then management is not eliminated but fundamentally reconfigured. PIS provides a framework for studying this transformation as triadic structuration: new divisions of labor between humans and AI (P↔I) produce new organizational structures (I↔S) that in turn reshape managerial roles and identities (P↔S).

*Cross-level dynamics.* PIS's multi-level architecture supports investigation of how micro-level human–AI interactions aggregate into meso-level organizational change and macro-level institutional transformation. For instance, individual physicians' appropriation of clinical AI (micro) may reshape hospital governance structures (meso) and ultimately transform medical education standards and regulatory frameworks (macro). Existing frameworks lack the multi-level apparatus to trace these cascading dynamics.

## 6.4 Implications for Practice

For organizational leaders and technology managers, PIS offers a diagnostic framework. When technology implementations fail or produce unintended consequences, PIS directs attention to triadic misalignment rather than single-point failures. A common pattern is to blame technology ("the system doesn't work") or people ("users resist change") when the actual source of dysfunction is misalignment among all three elements—for instance, an AI system that technically performs well (I) but undermines existing authority relations (S) in ways that people (P) experience as threatening.

PIS also suggests that successful AI implementation requires *triadic design*—simultaneous attention to the technology's capabilities, the organizational structures that govern its use, and the human practices through which it is appropriated. This goes beyond the STS principle of joint optimization (which addresses only two elements) to insist on three-way alignment.

Specifically, PIS implies several practical design principles. First, *structural co-design*: when implementing new IT (especially AI), organizations should simultaneously redesign governance structures, accountability mechanisms, and role definitions rather than treating these as post-implementation adjustments. The common practice of deploying technology first and "managing change" afterward violates the structuration principle that all three elements are co-constituted from the outset. Second, *appropriation awareness*: technology designers and implementers should expect and design for divergent appropriation. Since People will inevitably develop technologies-in-practice that differ from designed functionality, systems should be flexible enough to accommodate emergent uses while robust enough to maintain organizational coherence. Third, *tempo management*: given that the three structuration circuits operate at different tempos (Proposition 6), organizations should manage the pace of AI-driven change to prevent destructive misalignment. Rapid technological deployment that outpaces structural adaptation creates governance vacuums; institutional inertia that lags behind technological possibility creates missed opportunities and workaround cultures.

### 6.5 Boundary Conditions and Scope

The PIS framework is intended as a general theoretical architecture for MIS research, but its applicability varies across empirical contexts. Several boundary conditions merit discussion.

First, PIS is most analytically powerful in contexts where all three elements are actively in play—where People exercise meaningful agency, IT has substantive material properties that shape practice, and organizational Structures are neither so rigid as to prevent structuration nor so fluid as to provide no structural constraint. In highly routinized, tightly controlled environments (e.g., fully automated manufacturing lines), the People element may be minimally present; in purely informal, pre-institutional settings (e.g., early-stage startups with no formal structures), the Structure element may be underdeveloped. PIS remains applicable in these contexts but may require supplementation by domain-specific theories.

Second, PIS's structuration mechanism assumes ongoing recursive interaction over time. It is therefore better suited to studying processes of technology implementation, organizational transformation, and institutional evolution than to studying one-time events or cross-sectional states. Researchers seeking to explain point-in-time outcomes (e.g., adoption rates at a particular moment) may find variance models more parsimonious, though PIS would argue that such outcomes are themselves products of structuration processes that cross-sectional methods can only partially capture.

Third, the framework's critical realist ontology positions it between positivist and interpretivist traditions. Researchers committed to strong constructivism may find PIS's insistence on analytically distinct elements (particularly IT as materially real) too realist; those committed to positivist prediction may find PIS's recursive causation difficult to operationalize. We view this positioning as a strength—PIS occupies the "middle ground" that Mingers (2001) has long advocated for IS research—but acknowledge that it may not satisfy those seeking ontological purity.

## 7. Conclusion

The PIS (People–IT–Structuration) framework proposed in this paper represents an attempt to synthesize decades of MIS theorizing into a coherent, integrative framework. By identifying People, IT, and Structure as three analytically distinct but ontologically co-constitutive elements, and by specifying triadic structuration as the mechanism through which they interact, PIS addresses persistent gaps in the discipline's theoretical infrastructure.

The framework is not intended to replace existing theories but to provide a meta-theoretical architecture within which their insights can be integrated and extended. TAM's focus on individual perceptions, AST's attention to appropriation processes, sociomateriality's emphasis on constitutive entanglement—each captures genuine features of the P–I–S nexus. PIS's contribution is to show how these partial views fit together and to identify the triadic dynamics that none of them individually addresses.

The urgency of this integrative project is amplified by artificial intelligence. AI is not merely a new technology; it is a qualitative shift in the IT element that intensifies all three structuration circuits and generates novel phenomena—algorithmic structuration, human–AI collaboration, mediated agency—that existing frameworks cannot adequately theorize. PIS provides the theoretical vocabulary and analytical architecture for engaging with these phenomena systematically.

Several limitations should be acknowledged. First, as a conceptual framework, PIS requires empirical validation. The propositions advanced here are intended to be generative rather than directly testable; operationalizing them for specific empirical contexts will require careful specification of boundary conditions, measurement strategies, and contextual contingencies. Second, PIS inherits the challenges of structuration theory itself, particularly the difficulty of studying recursive processes using methods designed for linear causation. We have argued for methodological pluralism, but the practical challenges of conducting triadic structuration research should not be underestimated. Third, PIS's scope is ambitious—spanning micro-level interaction to macro-level institutional dynamics—and there is a risk that such breadth may come at the cost of precision in particular empirical applications. We encourage researchers to develop mid-range theories within the PIS architecture that specify mechanisms for particular domains, levels, or phenomena.

Despite these limitations, we believe the PIS framework addresses a genuine and consequential gap in MIS theorizing. The field's theoretical infrastructure was developed for an era of passive tools and relatively stable organizational arrangements. The AI era demands frameworks that can accommodate technology with agency-like properties, rapid organizational restructuring, and complex human–machine collaboration. PIS provides such a framework while remaining grounded in the discipline's richest theoretical traditions.

We invite the IS research community to engage with the PIS framework critically—testing its propositions empirically, extending its mechanisms theoretically, and challenging its assumptions where they fall short. The framework is offered not as a final answer to the discipline's theoretical challenges but as a generative starting point for a new phase of integrative theorizing in the AI era.